\documentclass[twocolumn, showpacs,preprintnumbers,amsmath,amssymb]{revtex4}
\usepackage{amssymb}
\usepackage{amsfonts}

\usepackage{graphicx}
\usepackage{dcolumn}
\usepackage{bm}

\begin{document}

\title{Nested Quantum Error Correction Codes}

\author{Zhuo Wang$^1$, Kai Sun$^2$}%
\author{Heng Fan$^1$\footnote{hfan@aphy.iphy.ac.cn}}
\author{Vlatko Vedral$^{3,4,5}$\footnote{vlatko.vedral@qubit.org}}
\affiliation{%
$^1$Institute of Physics, Chinese Academy of Sciences, Beijing
100190, China\\
$^2$Institute of Computing Technology, Chinese Academy of Sciences,
Beijing
100190, China \\
$^3$Clarendon Laboratory, University of Oxford, Parks Road, Oxford OX1 3PU, United Kingdom \\
$^4$Centre for Quantum Technologies, National University of
Singapore, 3
Science Drive 2, Singapore 117543 \\
$^5$Department of Physics, National University of Singapore, 2
Science Drive 3, Singapore 117542
}%
\date{\today}



\begin{abstract}
 The theory of quantum error correction was established more than a
decade ago as the primary tool for fighting decoherence in quantum
information processing \cite{shor,CS,steane,gottesman0}. Although
great progress has already been made in this field, limited methods
are available in constructing new quantum error correction codes
from old codes. Here we exhibit a simple and general method to
construct new quantum error correction codes by nesting certain
quantum codes together. The problem of finding long quantum error
correction codes is reduced to that of searching several short
length quantum codes with certain properties. Our method works for
all length and all distance codes, and is quite efficient to
construct optimal or near optimal codes. Two main known methods in
constructing new codes from old codes in quantum error-correction
theory, the concatenating and pasting, can be understood in the
framework of nested quantum error correction codes.
\end{abstract}
\maketitle

Quantum computers offer a means of solving certain problems,
including prime factorization, exponentially faster than classical
computers \cite{shor0}. One of the main difficulties in building the
quantum computers is that the quantum information are fragile to
decoherence. To solve this problem, the quantum error correction
codes (QECC) are designed which provide us an active way of
protecting our precious quantum data from quantum noises. Since the
initial discovery of quantum error-correcting coeds (QECC),
researchers have made great progress in codes construction. But for
large number of qubits there has been less progress, and only a few
general code constructions are know. The main difficulty in
constructing long codes is that we need to choose the suitable QECC
from a large quantity of quantum states whose number in general
grows exponentially with the length. This random search method will
be intractable for long codes. However, long QECC is indeed
necessary when we can control more and more qubits to realize the
scalable quantum computation. If we can divide the searching of QECC
into several easier steps, then the difficulty in constructing long
QECC will be reduced dramatically. In this work, we propose a
general and easy method to construct the {\it nested} QECC.

\section{Codes Construction for Distance 3}

\subsection{An optimal $[10,4,3]$ QECC}
Our main idea to construct long QECC from known short length QECC is
to ensure that all syndromes for one-qubit error of the long QECC
are different. For example, we may consider to repeat a $[5,1,3]$
QECC twice, automatically, the syndromes are also repeated twice.
Since the syndromes for $[5,1,3]$ are different, the next step to
construct a long QECC is to find additional generators so that the
syndromes in each block can be distinguished. So we consider to put
a code of length 2 whose syndromes are also different under any
block to achieve it. We can find the additional generators that take
the following form,
\begin{equation}
\left(
\begin{array}{cc}
  $1$ & $0$
\\$1$ & $1$
\end{array}
\right| \left.
\begin{array}{cc}
  $1$ & $1$
\\$0$ & $1$
\end{array}
\right).
\end{equation}
We name it subcode $\{2,2\}$, i.e., the length is two and the code
has two generators. To ensure that it is a valid QECC, we also need
to confirm that each generator of the above matrix, $(YZ)$ and
$(XY)$, which will be repeated 5 times in the new code commute with
each other and also commute with the original generators of
$[5,1,3]$ which is named blockcode. Suppose in $GF(2)$ each
generator of subcode is written as $(a|b)=(a_1a_2|b_1b_2)$ and for
blockcode as $(c|d)=(c_1c_2c_3c_4c_5|d_1d_2d_3d_4d_5)$, then
commuting condition for this construction will be
\begin{eqnarray}
&&(a_1+a_2)(d_1+d_2+d_3+d_4+d_5)\nonumber\\
&&+(b_1+b_2)(c_1+c_2+c_3+c_4+c_5)=0
\end{eqnarray}
(see (\ref{derive1}) and (\ref{derive2}) for details about how to
obtain this equation). However, we may consider a stronger
constraint condition, for example, $(c_1+c_2+c_3+c_4+c_5)=0$ and
$(d_1+d_2+d_3+d_4+d_5)=0$ which is for the blockcode. Really we can
find a $[5,1,3]$ code that satisfies this condition and the code
takes the form
\begin{equation}
\left(
\begin{array}{ccccc}
  $1$ & $0$ & $0$ & $1$ & $0$
\\$0$ & $1$ & $0$ & $0$ & $1$
\\$1$ & $0$ & $1$ & $0$ & $0$
\\$0$ & $1$ & $0$ & $1$ & $0$
\end{array}
\right| \left.
\begin{array}{ccccc}
  $0$ & $1$ & $1$ & $0$ & $0$
\\$0$ & $0$ & $1$ & $1$ & $0$
\\$0$ & $0$ & $0$ & $1$ & $1$
\\$1$ & $0$ & $0$ & $0$ & $1$
\end{array}
\right).
\end{equation}
So stabilizer codes constructed from $[5,1,3]$ and $\{2,2\}$
presented in the following constitute a new QECC.
\begin{eqnarray}
\left(
\begin{array}{l}
{\left. \begin{array}{l} {\begin{array}{cc} X&X\\
I&I\\
X&X\\
Z&Z\\
\end{array}}
\\
\overline{\begin{array}{cc} Y&Z\\
X&Y\end{array}}
\end{array}
\right|} {\left.
\begin{array}{l} {\begin{array}{cc} Z&Z\\
X&X\\
I&I\\
X&X\\
\end{array}}
\\
\overline{\begin{array}{cc} Y&Z\\
X&Y\end{array}}
\end{array}
\right|} {\left.
\begin{array}{l} {\begin{array}{cc} Z&Z\\
Z&Z\\
X&X\\
I&I\\
\end{array}}
\\
\overline{\begin{array}{cc} Y&Z\\
X&Y\end{array}}
\end{array}
\right|} {\left.
\begin{array}{l} {\begin{array}{cc} X&X\\
Z&Z\\
Z&Z\\
X&X\\
\end{array}}
\\
\overline{\begin{array}{cc} Y&Z\\
X&Y\end{array}}
\end{array}
\right|} {
\begin{array}{l} {\begin{array}{cc} I&I\\
X&X\\
Z&Z\\
Z&Z\\
\end{array}}
\\
\overline{\begin{array}{cc} Y&Z\\
X&Y\end{array}}
\end{array}
}
\end{array}
\right)
\end{eqnarray}
This is a $[10,4,3]$ code and we know it is optimal.

\subsection{Other optimal QECC constructed from $[5,1,3]$}
We can also find subcodes $\{3,2\}$ and $\{4,3\}$
\begin{equation}
\left(
\begin{array}{ccc}
  $1$ & $0$ & $0$
\\$1$ & $1$ & $0$
\end{array}
\right| \left.
\begin{array}{ccc}
  $1$ & $1$ & $0$
\\$0$ & $1$ & $0$
\end{array}
\right),
\end{equation}
\begin{equation}
\left(
\begin{array}{cccc}
  $1$ & $0$ & $0$ & $1$
\\$0$ & $1$ & $0$ & $1$
\\$0$ & $0$ & $1$ & $1$
\end{array}
\right| \left.
\begin{array}{cccc}
  $0$ & $0$ & $1$ & $0$
\\$1$ & $0$ & $1$ & $1$
\\$0$ & $1$ & $0$ & $1$
\end{array}
\right).
\end{equation}
We consider the same stronger constraint condition for blockcode.
From $[5,1,3]$ code, the new QECCs constructed are $[15,9,3]$ and
$[20,13,3]$ which are all optimal. Their explicit forms are
presented in the appendix (see (\ref{code15}) and (\ref{code20})).

\subsection{QECC of length 25: nesting two [5,1,3] codes}
A straightforward method for a length 25 codes from [5,1,3] code is
the concatenated QECC \cite{gottesman0}. Here our method is quite
different from the concatenation. To construct a QECC for $n=25$
from code $[5,1,3]$, a simple method is to let both the block-code
and the subcode be a $[5,1,3]$ code. Then commuting condition for
this construction will be
\begin{eqnarray}
&&(a_1+a_2+a_3+a_4+a_5)(d_1+d_2+d_3+d_4+d_5)\nonumber\\
&&+(b_1+b_2+b_3+b_4+b_5)(c_1+c_2+c_3+c_4+c_5)=0. \nonumber \\
\end{eqnarray}
However, we may consider a stronger constraint condition as
$(a_1+a_2+a_3+a_4+a_5)=0$ and $(c_1+c_2+c_3+c_4+c_5)=0$. Then we can
find a $[5,1,3]$ code for both blockcode and subcode that satisfies
this condition and the code takes the form
\begin{equation}
\left(
\begin{array}{ccccc}
  $1$ & $0$ & $0$ & $0$ & $1$
\\$0$ & $1$ & $0$ & $0$ & $1$
\\$0$ & $0$ & $1$ & $0$ & $1$
\\$0$ & $0$ & $0$ & $1$ & $1$
\end{array}
\right| \left.
\begin{array}{ccccc}
  $0$ & $0$ & $1$ & $1$ & $1$
\\$1$ & $0$ & $0$ & $1$ & $0$
\\$1$ & $1$ & $1$ & $0$ & $1$
\\$0$ & $1$ & $1$ & $1$ & $0$
\end{array}
\right).
\end{equation}

The newly constructed QECC is a $[25,17,3]$ code, see
(\ref{code2517}) which is not optimal. To construct the optimal code
for 25-qubit which should be a $[25,18,3]$ code, we need a $\{5,3\}$
code to be the subcode. Still we use the stronger constraint
condition $(c_1+c_2+c_3+c_4+c_5)=0$ and $(d_1+d_2+d_3+d_4+d_5)=0$
for the blockcode, then there will be no extra conditions for the
subcode, so any $\{5,3\}$ code can be used. One example of an
optimal $[25,18,3]$ code constructed is presented in the appendix
,see (\ref{code2518}).

We may notice that that syndrome $(0000)$ of code $[5,1,3]$ which is
not used in constructing code $[25,17,3]$. With this property in
mind, we can repeat this syndrome 5 times and then nest a $[5,1,3]$
as the subcode to this syndrome. Then a code $[30,22,3]$, see
(\ref{code30}) can be constructed, this is a near optimal QECC.
Continuously we can construct a $[35,27,3]$ code, see (\ref{code35})
which is also near optimal.

\subsection{Method to constructing perfect QECC}

Here we present our nested method in constructing one class of
codes, perfect codes \cite{gottesman0}, which are optimal since it
achieves the Hamming bound.

First we need a class of sub-codes, we name them as raw
perfect-constructing codes, which are one kind of $\{2^k-1,k\}$ QECC
whose all $(2^k-1)$ one-qubit $\sigma_x$ errors take all the $2^k-1$
syndromes that the code can provide except for syndrome
$(00\cdots0)$, and so do the $\sigma_z$ and $\sigma_y$ errors.
Suppose any generator of a $\{2^k-1,k\}$ raw perfect-constructing
code is $(c_1,\cdots,c_{2^k-1}|d_1,\cdots,d_{2^k-1})$, one feature
of those codes are $(c_1+\cdots+c_{2^k-1})=0$ and
$(d_1+\cdots+d_{2^k-1})=0$. So if we choose a raw
perfect-constructing code as subcode, all commuting conditions
satisfied automatically. So there will be no extra conditions for
the blockcode which can make the code searching much easier.

 We have the following results:

(1), A $[2^k,2^k-k-2,3]$ code can be constructed by nesting a "code"
$\left(
\begin{array}{c}
  $1$
\\$0$
\end{array}
\right| \left.
\begin{array}{c}
  $0$
\\$1$
\end{array}
\right)$ as blockcode and a $[2^k-1,2^k-1-k,<3]$ raw
perfect-constructing code with syndrome $(00\cdots0)$ as subcode
together.

(2), A $[(2^{k+k'}-1)/3,(2^{k+k'}-1)/3-(k+k'),3]$ perfect code can
be constructed by nesting a $[(2^k-1)/3,(2^k-1)/3-k,3]$ perfect
code, a $[2^{k'}-1,2^{k'}-1-{k'},<3]$ raw perfect-constructing code
and a $[(2^{k'}-1)/3,(2^{k'}-1)/3-{k'},3]$ perfect code together.

We take the constructing of $[16,10,3]$ code perfect code as
examples and present the detail in appendix, see (\ref{code16}).

Gottesman's stabilizer pasting of distance 3 code is also easy to be
understood in our theory, but our method is more general for two
reasons: (i) blockcode of $[n_2,s_2]$ can be any code not only
$\left(
\begin{array}{c}
  $1$
\\$0$
\end{array}
\right| \left.
\begin{array}{c}
  $0$
\\$1$
\end{array}
\right)$, which has been shown in constructing $[30,22,3]$ code
 and $[35,27,3]$ code and perfect code presented in above;
 (ii) the syndromes of blockcode
 of $[n_1,s_1]$ need not to be $00\cdots0$ only. In Ref.\cite{yu}, Yu {\it et al.} present a
 construction of optimal $[37,30,3]$ code which is $[37]=[2^5]\rhd[5]$,
 here we give another construction which is more powerful,
 the detail can be found in appendix, see (\ref{code37}).

\section{Nesting Codes for all Distance}
Not only for distance 3, our method works for all distance QECC.
Suppose there are two copies of a subcode whose generator matrix is
$[A]$, the easiest way to distinguish their syndromes
is to connect them in block diagonal matrix as $\bordermatrix{%
& & \cr
 &A &0\cr
 &0 &A\cr}$ since the logical operators of this construction are only
 constructed by logical operators of these two
 codes,
and the hardest way is to connect them parallel which is$\bordermatrix{%
& & \cr
 &A &A\cr}$. For the hardest way, the problem of constructing
 distance 3 codes have been solved above, and problem of constructing larger
 distance codes is still open. But for the easiest way, codes
 construction for all distance can be solved which will be presented in
 below.

Suppose there are $m$ subcodes, i.e., $[n_1,k_1,d_1]$,
$[n_2,k_2,d_2]$, $\cdots$, $[n_m,k_m,d_m]$, that are connected in
block diagonal matrix whose logical operators are
$\{\bar{X_1^1},\cdots,\bar{X_{k_1}^1},\bar{Z_1^1},\cdots,\bar{Z_{k_1}^1}\}$,
$\cdots$,
$\{\bar{X_1^m},\cdots,\bar{X_{k_m}^m},\bar{Z_1^m},\cdots,\bar{Z_{k_m}^m}\}$.
The blockcode is a $n=\sum^m_{i=1}k_i$ code whose weight of elements
should be redefined: as long as there has at least one $\sigma_x$ or
$\sigma_z$ or $\sigma_y$ in the first $k_1$ qubits we count "1", and
the same to next $k_2$, $K_3$, $\cdots$, $k_m$ qubits. We nest
blockcode and subcode as the following: any physical qubit of
blockcode is replaced by the corresponding logical operator of
subcode, then if we could find a redefined $[\sum^m_{i=1}k_i,k',d']$
code and if $d$ is defined as the minimum sum of $d'$ numbers of
$\{d_1,\cdots,d_m\}$, a new $[\sum^m_{i=1}n_i,k',\geq d]$ degenerate
QECC is constructed.

The details will be presented in appendix. Our theory are general
and for some special cases, it reduces to some known methods of
constructing codes:  (1) if all subcodes are the same codes and
$k=1$, it will turn to the theory of concatenated coding (see
\ref{concatenatedcode}); (2) if the number of subcode is 2, it will
turn to the theory of general stabilizer pasting for all distance .

\newpage
\widetext

\appendix
\section{\label{sec:level1}Codes Construction for Distance 3}

An $[n,n-k,3]$ stabilizer code has k generators which build a
$(k\times 2n)$ generator matrix. There are $3n$ one-qubit error
syndromes in which any bit flip error syndrome $f(\sigma_x^i)$ is
the $i$th column of the right-half matrix, and any sigh flip error
syndrome $f(\sigma_z^i)$ is the $i$th column of the left-half matrix
and $f(\sigma_y^i)$ is the sum of these two columns. If the code is
nondegenerate, then all the $3n$ syndromes are different.

 For a $[n,n-k,d]$ code to correct $t$ errors, we define the using rate of
syndromes:
\begin{eqnarray}
g(n,k,t)=\frac{\sum_{j=1}^t3^jC_n^j}{2^k}, \label{model}
\end{eqnarray}
which is the division of the syndromes that all errors used and the
syndromes that the code can provide. Using rate of syndromes is a
very important parameter in codes construction, and obviously when
$g(n,k,t)\leq1$, the larger $g$ is, the more optimal the code is.
For a distance 3 nondegenerate code if $g=1-\frac{1}{2^k}$, we call
it perfect code.

 There are four steps to construct a $[n,n-k,3]$ nondegenerate
code: (i) choose $2n$ different syndromes for all $\sigma_x$ and
$\sigma_z$ errors; (ii) calculate $\sigma_y$ syndromes and make sure
that all $3n$ syndromes are different; (iii) build the generator
matrix with $\sigma_x$ and $\sigma_z$ error syndromes and make sure
that all $k$ generators are commute to each other (we all it
commuting condition); (iv) make sure that all generators are real
generators which means no one is product of other generators (we
call it generator condition). Here we do not only restrict our focus
on real stabilizer code but also discuss the complex stabilizer
code, which means the number of $\sigma_y$ in each generator need
not to be even only.

 It is not easy to construct a code since step (iii) is a strong constraint condition and not easy to achieve. So
constructing an optimal or near optimal code whose using rate of
syndromes $g(n,k,t)$ is equal to or near one is much harder.

\subsection{Nesting two $[5,1,3]$ codes together}

$[5,1,3]$ is the first prefect distance 3 code and has many good
features that we need: (i) nondegenerate, which means all error
syndromes are different; (ii) perfect, $g(5,4,1)=15/16$, which means
it used all the syndromes that the code can provide except for 0000;
(iii) commuting, which means when we use $[5,1,3]$ codes to
construct new codes, some parts of the commuting condition of new
codes are satisfied automatically.

A $[25,17,3]$ code has 8 generators, which means the syndrome is a
8-bit number whose first 4-bit could define $2^4=16$ syndromes. It
is quite obvious that we can choose a $[5,1,3]$ code to take the
first 4-bit syndromes, and the using rate will be 15/16. Here we
call this $[5,1,3]$ code $blockcode$. So the blockcode which takes
the first 4-bit of the 8-bit syndromes defines 15 blocks, i.e.
$\{B_x^1,B_x^2,B_x^3,B_x^4,B_x^5,B_z^1,B_z^2,B_z^3,B_z^4,B_z^5,B_y^1,B_y^2,B_y^3,B_y^4,B_y^5\}$,
which are called $blockcode$ syndromes and satisfy:
\begin{eqnarray}
B_x^i+B_z^i=B_y^i
\end{eqnarray}
and make all blocks different from each other.

Then what we need to do is only to make the last 4-bit syndromes in
any block different and make sure that any $\sigma_y$ syndrome is
the sum of corresponding $\sigma_x$ and $\sigma_z$ syndromes, then
step (i) and (ii) of code construction will be finished. We can see
that also a $[5,1,3]$ code could satisfy this condition. We call
this code $subcode$ and the last 4-bit syndromes subcode syndromes,
which is
$\{S_x^1,S_x^2,S_x^3,S_x^4,S_x^5,S_z^1,S_z^2,S_z^3,S_z^4,S_z^5,S_y^1,S_y^2,S_y^3,S_y^4,S_y^5\}$
and
\begin{eqnarray}
S_x^i+S_z^i=S_y^i
\end{eqnarray}

So blockcode and subcode are nested in this way:
\begin{eqnarray}
\{\{B_x^1,B_x^2,B_x^3,B_x^4,B_x^5\}\otimes\{S_x^1,S_x^2,S_x^3,S_x^4,S_x^5\},\nonumber\\
\{B_z^1,B_z^2,B_z^3,B_z^4,B_z^5\}\otimes\{S_z^1,S_z^2,S_z^3,S_z^4,S_z^5\},\nonumber\\
\{B_y^1,B_y^2,B_y^3,B_y^4,B_y^5\}\otimes\{S_y^1,S_y^2,S_y^3,S_y^4,S_y^5\}\},
\end{eqnarray}
which means if the $[5,1,3]$ blockcode and subcode are
\begin{equation}
\left(
\begin{array}{ccccc}
  a_{11} & a_{12} & a_{13} & a_{14} & a_{15}
\\a_{21} & a_{22} & a_{23} & a_{24} & a_{25}
\\a_{31} & a_{32} & a_{33} & a_{34} & a_{35}
\\a_{41} & a_{42} & a_{43} & a_{44} & a_{45}
\end{array}
\right| \left.
\begin{array}{ccccc}
  b_{11} & b_{12} & b_{13} & b_{14} & b_{15}
\\b_{21} & b_{22} & b_{23} & b_{24} & b_{25}
\\b_{31} & b_{32} & b_{33} & b_{34} & b_{35}
\\b_{41} & b_{42} & b_{43} & b_{44} & b_{45}
\end{array}
\right)
\end{equation}
\begin{equation}
and \left(
\begin{array}{ccccc}
  c_{11} & c_{12} & c_{13} & c_{14} & c_{15}
\\c_{21} & c_{22} & c_{23} & c_{24} & c_{25}
\\c_{31} & c_{32} & c_{33} & c_{34} & c_{35}
\\c_{41} & c_{42} & c_{43} & c_{44} & c_{45}
\end{array}
\right| \left.
\begin{array}{ccccc}
  d_{11} & d_{12} & d_{13} & d_{14} & d_{15}
\\d_{21} & d_{22} & d_{23} & d_{24} & d_{25}
\\d_{31} & d_{32} & d_{33} & d_{34} & d_{35}
\\d_{41} & d_{42} & d_{43} & d_{44} & d_{45}
\end{array},
\right)
\end{equation}
the new nested code will be
{\footnotesize
\begin{eqnarray}
\begin{array}{l}
{\left(
\begin{array}{ccccccccccccccccccccccccc}
  a_{11} & a_{11} & a_{11} & a_{11} & a_{11} & a_{12} & a_{12} & a_{12} & a_{12} & a_{12} & a_{13} & a_{13} & a_{13} & a_{13} & a_{13} & a_{14} & a_{14} & a_{14} & a_{14} & a_{14} & a_{15} & a_{15} & a_{15} & a_{15} & a_{15}
\\a_{21} & a_{21} & a_{21} & a_{21} & a_{21} & a_{22} & a_{22} & a_{22} & a_{22} & a_{22} & a_{23} & a_{23} & a_{23} & a_{23} & a_{23} & a_{24} & a_{24} & a_{24} & a_{24} & a_{24} & a_{25} & a_{25} & a_{25} & a_{25} & a_{25}
\\a_{31} & a_{31} & a_{31} & a_{31} & a_{31} & a_{32} & a_{32} & a_{32} & a_{32} & a_{32} & a_{33} & a_{33} & a_{33} & a_{33} & a_{33} & a_{34} & a_{34} & a_{34} & a_{34} & a_{34} & a_{35} & a_{35} & a_{35} & a_{35} & a_{35}
\\a_{41} & a_{41} & a_{41} & a_{41} & a_{41} & a_{42} & a_{42} & a_{42} & a_{42} & a_{42} & a_{43} & a_{43} & a_{43} & a_{43} & a_{43} & a_{44} & a_{44} & a_{44} & a_{44} & a_{44} & a_{45} & a_{45} & a_{45} & a_{45} & a_{45}
\\c_{11} & c_{12} & c_{13} & c_{14} & c_{15} & c_{11} & c_{12} & c_{13} & c_{14} & c_{15} & c_{11} & c_{12} & c_{13} & c_{14} & c_{15} & c_{11} & c_{12} & c_{13} & c_{14} & c_{15} & c_{11} & c_{12} & c_{13} & c_{14} & c_{15}
\\c_{21} & c_{22} & c_{23} & c_{24} & c_{25} & c_{21} & c_{22} & c_{23} & c_{24} & c_{25} & c_{21} & c_{22} & c_{23} & c_{24} & c_{25} & c_{21} & c_{22} & c_{23} & c_{24} & c_{25} & c_{21} & c_{22} & c_{23} & c_{24} & c_{25}
\\c_{31} & c_{32} & c_{33} & c_{34} & c_{35} & c_{31} & c_{32} & c_{33} & c_{34} & c_{35} & c_{31} & c_{32} & c_{33} & c_{34} & c_{35} & c_{31} & c_{32} & c_{33} & c_{34} & c_{35} & c_{31} & c_{32} & c_{33} & c_{34} & c_{35}
\\c_{41} & c_{42} & c_{43} & c_{44} & c_{45} & c_{41} & c_{42} & c_{43} & c_{44} & c_{45} & c_{41} & c_{42} & c_{43} & c_{44} & c_{45} & c_{41} & c_{42} & c_{43} & c_{44} & c_{45} & c_{41} & c_{42} & c_{43} & c_{44} & c_{45}
\end{array}
\right| }
\\
{\left| \begin{array}{ccccccccccccccccccccccccc}
  b_{11} & b_{11} & b_{11} & b_{11} & b_{11} & b_{12} & b_{12} & b_{12} & b_{12} & b_{12} & b_{13} & b_{13} & b_{13} & b_{13} & b_{13} & b_{14} & b_{14} & b_{14} & b_{14} & b_{14} & b_{15} & b_{15} & b_{15} & b_{15} & b_{15}
\\b_{21} & b_{21} & b_{21} & b_{21} & b_{21} & b_{22} & b_{22} & b_{22} & b_{22} & b_{22} & b_{23} & b_{23} & b_{23} & b_{23} & b_{23} & b_{24} & b_{24} & b_{24} & b_{24} & b_{24} & b_{25} & b_{25} & b_{25} & b_{25} & b_{25}
\\b_{31} & b_{31} & b_{31} & b_{31} & b_{31} & b_{32} & b_{32} & b_{32} & b_{32} & b_{32} & b_{33} & b_{33} & b_{33} & b_{33} & b_{33} & b_{34} & b_{34} & b_{34} & b_{34} & b_{34} & b_{35} & b_{35} & b_{35} & b_{35} & b_{35}
\\b_{41} & b_{41} & b_{41} & b_{41} & b_{41} & b_{42} & b_{42} & b_{42} & b_{42} & b_{42} & b_{43} & b_{43} & b_{43} & b_{43} & b_{43} & b_{44} & b_{44} & b_{44} & b_{44} & b_{44} & b_{45} & b_{45} & b_{45} & b_{45} & b_{45}
\\d_{11} & d_{12} & d_{13} & d_{14} & d_{15} & d_{11} & d_{12} & d_{13} & d_{14} & d_{15} & d_{11} & d_{12} & d_{13} & d_{14} & d_{15} & d_{11} & d_{12} & d_{13} & d_{14} & d_{15} & d_{11} & d_{12} & d_{13} & d_{14} & d_{15}
\\d_{21} & d_{22} & d_{23} & d_{24} & d_{25} & d_{21} & d_{22} & d_{23} & d_{24} & d_{25} & d_{21} & d_{22} & d_{23} & d_{24} & d_{25} & d_{21} & d_{22} & d_{23} & d_{24} & d_{25} & d_{21} & d_{22} & d_{23} & d_{24} & d_{25}
\\d_{31} & d_{32} & d_{33} & d_{34} & d_{35} & d_{31} & d_{32} & d_{33} & d_{34} & d_{35} & d_{31} & d_{32} & d_{33} & d_{34} & d_{35} & d_{31} & d_{32} & d_{33} & d_{34} & d_{35} & d_{31} & d_{32} & d_{33} & d_{34} & d_{35}
\\d_{41} & d_{42} & d_{43} & d_{44} & d_{45} & d_{41} & d_{42} & d_{43} & d_{44} & d_{45} & d_{41} & d_{42} & d_{43} & d_{44} & d_{45} & d_{41} & d_{42} & d_{43} & d_{44} & d_{45} & d_{41} & d_{42} & d_{43} & d_{44} & d_{45}
\end{array}
\right)}
\end{array}
\end{eqnarray}
}

Step (iii) and (iv) are usually the hardest part for code
construction, but in our method it is much easier to achieve. No
matter how many times blockcode and subcode repeat, commuting
conditions of themselves are also satisfied, so what we need to do
is to satisfy commuting condition between blockcode and subcode
which is
\begin{eqnarray}
(a_{i1}+a_{i2}+a_{i3}+a_{i4}+a_{i5})(d_{j1}+d_{j2}+d_{j3}+d_{j4}+d_{j5})\nonumber\\
+(b_{i1}+b_{i2}+b_{i3}+b_{i4}+b_{i5})(c_{j1}+c_{j2}+c_{j3}+c_{j4}+c_{j5})=0
\end{eqnarray}

then we can consider a more stronger constraint condition which is
$(a_{i1}+a_{i2}+a_{i3}+a_{i4}+a_{i5})=0$ and
$(c_{j1}+c_{j2}+c_{j3}+c_{j4}+c_{j5})=0$ that means the number of
"1"s in any row of left halves of generator matrices of both
blockcode and subcode is even.

For step (iv), obviously blockcode and subcode satisfy generator
condition themselves. It is easy to prove that between blockcode and
subcode the only opportunity that one generator is product of others
is $(00\cdots0|11\cdots1)$ or $(11\cdots1|00\cdots0)$ or
$(11\cdots1|11\cdots1)$ belongs to stabilizer of both codes at the
same time, and the probability is quite small. For computer search
Then we turn to find codes that all $(00\cdots0|11\cdots1)$,
$(11\cdots1|00\cdots0)$ and $(11\cdots1|11\cdots1)$ do not belong to
$S$ which is a more stronger constraint condition for computer
search.

Depending on the computer search, we find many accordant $[5,1,3]$
code, one of which is presented in the article, then the $[25,17,3]$
new code by nesting two $[5,1,3]$ codes is
\begin{equation}\label{code2517}
\left(
\begin{array}{ccccccccccccccccccccccccc}
  $X$ & $X$ & $X$ & $X$ & $X$ & $I$ & $I$ & $I$ & $I$ & $I$ & $Z$ & $Z$ & $Z$ & $Z$ & $Z$ & $Z$ & $Z$ & $Z$ & $Z$ & $Z$ & $Y$ & $Y$ & $Y$ & $Y$  &$Y$
\\$Z$ & $Z$ & $Z$ & $Z$ & $Z$ & $X$ & $X$ & $X$ & $X$ & $X$ & $I$ & $I$ & $I$ & $I$ & $I$ & $Z$ & $Z$ & $Z$ & $Z$ & $Z$ & $X$ & $X$ & $X$ & $X$  &$X$
\\$Z$ & $Z$ & $Z$ & $Z$ & $Z$ & $Z$ & $Z$ & $Z$ & $Z$ & $Z$ & $Y$ & $Y$ & $Y$ & $Y$  &$Y$ & $I$ & $I$ & $I$ & $I$ & $I$ & $Y$ & $Y$ & $Y$ & $Y$  &$Y$
\\$I$ & $I$ & $I$ & $I$ & $I$ & $Z$ & $Z$ & $Z$ & $Z$ & $Z$ & $Z$ & $Z$ & $Z$ & $Z$ & $Z$ & $Y$ & $Y$ & $Y$ & $Y$ & $Y$ & $X$ & $X$ & $X$ & $X$  &$X$
\\$X$ & $I$ & $Z$ & $Z$ & $Y$ & $X$ & $I$ & $Z$ & $Z$ & $Y$ & $X$ & $I$ & $Z$ & $Z$ & $Y$ & $X$ & $I$ & $Z$ & $Z$ & $Y$ & $X$ & $I$ & $Z$ & $Z$ & $Y$
\\$Z$ & $X$ & $I$ & $Z$ & $X$ & $Z$ & $X$ & $I$ & $Z$ & $X$ & $Z$ & $X$ & $I$ & $Z$ & $X$ & $Z$ & $X$ & $I$ & $Z$ & $X$ & $Z$ & $X$ & $I$ & $Z$ & $X$
\\$Z$ & $Z$ & $Y$ & $I$ & $Y$ & $Z$ & $Z$ & $Y$ & $I$ & $Y$ & $Z$ & $Z$ & $Y$ & $I$ & $Y$ & $Z$ & $Z$ & $Y$ & $I$ & $Y$ & $Z$ & $Z$ & $Y$ & $I$ & $Y$
\\$I$ & $Z$ & $I$ & $Y$ & $X$ & $I$ & $Z$ & $I$ & $Y$ & $X$ & $I$ & $Z$ & $I$ & $Y$ & $X$ & $I$ & $Z$ & $I$ & $Y$ & $X$ & $I$ & $Z$ & $I$ & $Y$ & $X$
\end{array}
\right).
\end{equation}

Here we discuss the using rate of syndromes in aspect of code
nesting. $g$ of both blockcode and subcode are 15/16, but each
blockcode syndrome only meets 1/3 of all subcode syndromes. So the
using rate is:
\begin{eqnarray}
g(25,8,1)=\frac{1}{3}g_{block}(5,4,1)g_{sub}(5,4,1)=\frac{75}{256}\leq\frac{1}{2}
\end{eqnarray}
which is the same as the result that calculated by equation (1). As
we know that the optimal 25-qubit code is $[25,18,3]$ whose using
rate is 75/128, this code is near optimal. So $g$ is an important
parameter to judge whether a code is optimal or not: generally if
$\frac{1}{2}<g\leq1$, the code must be optimal.

In the same way, we nest this $[5,1,3]$ code and this $[25,17,3]$
code to construct a $[125,113,3]$ code, and it is easy to prove that
the code satisfies all constructing conditions automatically. So all
the $[5^n,5^n-4n,3]$ codes could be constructed.

Then the $[30,22,3]$ near optimal code takes the form:
\begin{eqnarray}
\{\{B_x^1,B_x^2,B_x^3,B_x^4,B_x^5,0000\}\otimes\{S_x^1,S_x^2,S_x^3,S_x^4,S_x^5\},\nonumber\\
\{B_z^1,B_z^2,B_z^3,B_z^4,B_z^5,0000\}\otimes\{S_z^1,S_z^2,S_z^3,S_z^4,S_z^5\},\nonumber\\
\{B_y^1,B_y^2,B_y^3,B_y^4,B_y^5,0000\}\otimes\{S_y^1,S_y^2,S_y^3,S_y^4,S_y^5\}\},
\end{eqnarray}
so the generator matrix is
\begin{equation}\label{code30}
\left(
\begin{array}{cccccccccccccccccccccccccccccc}
  $X$ & $X$ & $X$ & $X$ & $X$ & $I$ & $I$ & $I$ & $I$ & $I$ & $Z$ & $Z$ & $Z$ & $Z$ & $Z$ & $Z$ & $Z$ & $Z$ & $Z$ & $Z$ & $Y$ & $Y$ & $Y$ & $Y$  &$Y$ & $I$ & $I$ & $I$ & $I$ & $I$
\\$Z$ & $Z$ & $Z$ & $Z$ & $Z$ & $X$ & $X$ & $X$ & $X$ & $X$ & $I$ & $I$ & $I$ & $I$ & $I$ & $Z$ & $Z$ & $Z$ & $Z$ & $Z$ & $X$ & $X$ & $X$ & $X$  &$X$ & $I$ & $I$ & $I$ & $I$ & $I$
\\$Z$ & $Z$ & $Z$ & $Z$ & $Z$ & $Z$ & $Z$ & $Z$ & $Z$ & $Z$ & $Y$ & $Y$ & $Y$ & $Y$  &$Y$ & $I$ & $I$ & $I$ & $I$ & $I$ & $Y$ & $Y$ & $Y$ & $Y$  &$Y$ & $I$ & $I$ & $I$ & $I$ & $I$
\\$I$ & $I$ & $I$ & $I$ & $I$ & $Z$ & $Z$ & $Z$ & $Z$ & $Z$ & $Z$ & $Z$ & $Z$ & $Z$ & $Z$ & $Y$ & $Y$ & $Y$ & $Y$ & $Y$ & $X$ & $X$ & $X$ & $X$  &$X$ & $I$ & $I$ & $I$ & $I$ & $I$
\\$X$ & $I$ & $Z$ & $Z$ & $Y$ & $X$ & $I$ & $Z$ & $Z$ & $Y$ & $X$ & $I$ & $Z$ & $Z$ & $Y$ & $X$ & $I$ & $Z$ & $Z$ & $Y$ & $X$ & $I$ & $Z$ & $Z$ & $Y$ & $X$ & $I$ & $Z$ & $Z$ & $Y$
\\$Z$ & $X$ & $I$ & $Z$ & $X$ & $Z$ & $X$ & $I$ & $Z$ & $X$ & $Z$ & $X$ & $I$ & $Z$ & $X$ & $Z$ & $X$ & $I$ & $Z$ & $X$ & $Z$ & $X$ & $I$ & $Z$ & $X$ & $Z$ & $X$ & $I$ & $Z$ & $X$
\\$Z$ & $Z$ & $Y$ & $I$ & $Y$ & $Z$ & $Z$ & $Y$ & $I$ & $Y$ & $Z$ & $Z$ & $Y$ & $I$ & $Y$ & $Z$ & $Z$ & $Y$ & $I$ & $Y$ & $Z$ & $Z$ & $Y$ & $I$ & $Y$ & $Z$ & $Z$ & $Y$ & $I$ & $Y$
\\$I$ & $Z$ & $I$ & $Y$ & $X$ & $I$ & $Z$ & $I$ & $Y$ & $X$ & $I$ & $Z$ & $I$ & $Y$ & $X$ & $I$ & $Z$ & $I$ & $Y$ & $X$ & $I$ & $Z$ & $I$ & $Y$ & $X$ & $I$ & $Z$ & $I$ & $Y$ & $X$
\end{array}
\right).
\end{equation}
Similarly all $[6\cdot5^n,6\cdot5^n-4(n+1),3]$ code and
$[5\cdot6^n,5\cdot6^n-4(n+1),3]$ codes could be constructed.

Then the $[35,27,3]$ code from the $[30,22,3]$ code takes the form:
\begin{eqnarray}
\{\{B_x^1,B_x^2,B_x^3,B_x^4,B_x^5,0000\}\otimes\{S_x^1,S_x^2,S_x^3,S_x^4,S_x^5\},\nonumber\\
\{B_x^1,B_x^2,B_x^3,B_x^4,B_x^5\}\otimes0000,\nonumber\\
\{B_z^1,B_z^2,B_z^3,B_z^4,B_z^5,0000\}\otimes\{S_z^1,S_z^2,S_z^3,S_z^4,S_z^5\},\nonumber\\
\{B_z^1,B_z^2,B_z^3,B_z^4,B_z^5\}\otimes0000,\nonumber\\
\{B_y^1,B_y^2,B_y^3,B_y^4,B_y^5,0000\}\otimes\{S_y^1,S_y^2,S_y^3,S_y^4,S_y^5\},\nonumber\\
\{B_y^1,B_y^2,B_y^3,B_y^4,B_y^5\}\otimes0000\}
\end{eqnarray}
whose generator matrix is
\begin{equation}\label{code35}
\left(
\begin{array}{ccccccccccccccccccccccccccccccccccc}
  $X$ & $X$ & $X$ & $X$ & $X$ & $X$ & $I$ & $I$ & $I$ & $I$ & $I$ & $I$ & $Z$ & $Z$ & $Z$ & $Z$ & $Z$ & $Z$ & $Z$ & $Z$ & $Z$ & $Z$ & $Z$ & $Z$ & $Y$ & $Y$ & $Y$ & $Y$  &$Y$ & $Y$ & $I$ & $I$ & $I$ & $I$ & $I$
\\$Z$ & $Z$ & $Z$ & $Z$ & $Z$ & $Z$ & $X$ & $X$ & $X$ & $X$ & $X$ & $X$ & $I$ & $I$ & $I$ & $I$ & $I$ & $I$ & $Z$ & $Z$ & $Z$ & $Z$ & $Z$ & $Z$ & $X$ & $X$ & $X$ & $X$  &$X$ & $X$ & $I$ & $I$ & $I$ & $I$ & $I$
\\$Z$ & $Z$ & $Z$ & $Z$ & $Z$ & $Z$ & $Z$ & $Z$ & $Z$ & $Z$ & $Z$ & $Z$ & $Y$ & $Y$ & $Y$ & $Y$ & $Y$ & $Y$ & $I$ & $I$ & $I$ & $I$ & $I$ & $I$ & $Y$ & $Y$ & $Y$ & $Y$  &$Y$ & $Y$ & $I$ & $I$ & $I$ & $I$ & $I$
\\$I$ & $I$ & $I$ & $I$ & $I$ & $I$ & $Z$ & $Z$ & $Z$ & $Z$ & $Z$ & $Z$ & $Z$ & $Z$ & $Z$ & $Z$ & $Z$ & $Z$ & $Y$ & $Y$ & $Y$ & $Y$ & $Y$ & $Y$ & $X$ & $X$ & $X$ & $X$  &$X$ & $X$ & $I$ & $I$ & $I$ & $I$ & $I$
\\$X$ & $I$ & $Z$ & $Z$ & $Y$ & $I$ & $X$ & $I$ & $Z$ & $Z$ & $Y$ & $I$ & $X$ & $I$ & $Z$ & $Z$ & $Y$ & $I$ & $X$ & $I$ & $Z$ & $Z$ & $Y$ & $I$ & $X$ & $I$ & $Z$ & $Z$ & $Y$ & $I$ & $X$ & $I$ & $Z$ & $Z$ & $Y$
\\$Z$ & $X$ & $I$ & $Z$ & $X$ & $I$ & $Z$ & $X$ & $I$ & $Z$ & $X$ & $I$ & $Z$ & $X$ & $I$ & $Z$ & $X$ & $I$ & $Z$ & $X$ & $I$ & $Z$ & $X$ & $I$ & $Z$ & $X$ & $I$ & $Z$ & $X$ & $I$ & $Z$ & $X$ & $I$ & $Z$ & $X$
\\$Z$ & $Z$ & $Y$ & $I$ & $Y$ & $I$ & $Z$ & $Z$ & $Y$ & $I$ & $Y$ & $I$ & $Z$ & $Z$ & $Y$ & $I$ & $Y$ & $I$ & $Z$ & $Z$ & $Y$ & $I$ & $Y$ & $I$ & $Z$ & $Z$ & $Y$ & $I$ & $Y$ & $I$ & $Z$ & $Z$ & $Y$ & $I$ & $Y$
\\$I$ & $Z$ & $I$ & $Y$ & $X$ & $I$ & $I$ & $Z$ & $I$ & $Y$ & $X$ & $I$ & $I$ & $Z$ & $I$ & $Y$ & $X$ & $I$ & $I$ & $Z$ & $I$ & $Y$ & $X$ & $I$ & $I$ & $Z$ & $I$ & $Y$ & $X$ & $I$ & $I$ & $Z$ & $I$ & $Y$ & $X$
\end{array}
\right).
\end{equation}

\subsection{Constructing optimal codes}

In our method to construct a distance 3 code, since each blockcode
syndrome only meets 1/3 of all subcode syndromes, the using rate
\begin{eqnarray}
g=\frac{1}{3}g_{block}g_{sub}
\end{eqnarray}
If a code is nested by two optimal codes, $g$ will always be less
than 1/3, which means the code will never be optimal. So if we want
to construct an optimal code, we need to find codes whose $g$ is
more than one to be blockcode or subcode. We call these codes whose
$g$ are more than one $over-optimal$ codes.

Suppose block and subcode are nested in this form:
\begin{eqnarray}
\{\{B_x^1,\cdots,B_x^n\}\otimes\{S_x^1,\cdots,S_x^{n'}\},\nonumber\\
\{B_z^1,\cdots,B_z^n\}\otimes\{S_z^1,\cdots,S_z^{n'}\},\nonumber\\
\{B_y^1,\cdots,B_y^n\}\otimes\{S_y^1,\cdots,S_y^{n'}\}\},
\end{eqnarray}
and suppose blockcode is an optimal distance 3 code, since each
$B_x$ syndrome only meets all $S_x$ syndromes and each $B_z$
syndrome only meets all $S_z$ syndromes and each $B_y$ syndrome only
meets all $S_y$ syndromes, there is no need to make any syndrome of
subcode be different, which means the distance od subcode need not
to be 3. We just need to find out a code whose $S_x$, $S_z$ and
$S_y$ syndromes are different themselves.

\subsubsection{the optimal $[10,4,3]$ QECC}

As we know, the optimal 10 qubits distance 3 code is $[10,4,3]$,
which means if we want to construct this code, the subcode must be a
$[2,0,<3]$ over-optimal code which we write as $\{2,2\}$. Suppose
blockcode and subcode are
\begin{equation}
\left(
\begin{array}{ccccc}
  c_{11} & c_{12} & c_{13} & c_{14} & c_{15}
\\c_{21} & c_{22} & c_{23} & c_{24} & c_{25}
\\c_{31} & c_{32} & c_{33} & c_{34} & c_{35}
\\c_{41} & c_{42} & c_{43} & c_{44} & c_{45}
\end{array}
\right| \left.
\begin{array}{ccccc}
  d_{11} & d_{12} & d_{13} & d_{14} & d_{15}
\\d_{21} & d_{22} & d_{23} & d_{24} & d_{25}
\\d_{31} & d_{32} & d_{33} & d_{34} & d_{35}
\\d_{41} & d_{42} & d_{43} & d_{44} & d_{45}
\end{array}
\right)
\end{equation}
\begin{equation}
and \left(
\begin{array}{cc}
  a_{11} & a_{12}
\\a_{21} & a_{22}
\end{array}
\right| \left.
\begin{array}{cc}
  b_{11} & b_{12}
\\b_{21} & b_{22}
\end{array},
\right)
\end{equation}
then the new nested code will be
\begin{equation}\label{derive1}
\left(
\begin{array}{cccccccccc}
  c_{11} & c_{11} & c_{12} & c_{12} & c_{13} & c_{13} & c_{14} & c_{14} & c_{15} & c_{15}
\\c_{21} & c_{21} & c_{22} & c_{22} & c_{23} & c_{23} & c_{24} & c_{24} & c_{25} & c_{25}
\\c_{31} & c_{31} & c_{32} & c_{32} & c_{33} & c_{33} & c_{34} & c_{34} & c_{35} & c_{35}
\\c_{41} & c_{41} & c_{42} & c_{42} & c_{43} & c_{43} & c_{44} & c_{44} & c_{45} & c_{45}
\\a_{11} & a_{12} & a_{11} & a_{12} & a_{11} & a_{12} & a_{11} & a_{12} & a_{11} & a_{12}
\\a_{21} & a_{22} & a_{21} & a_{22} & a_{21} & a_{22} & a_{21} & a_{22} & a_{21} & a_{22}
\end{array}
\right| \left.
\begin{array}{cccccccccc}
  d_{11} & d_{11} & d_{12} & d_{12} & d_{13} & d_{13} & d_{14} & d_{14} & d_{15} & d_{15}
\\d_{21} & d_{21} & d_{22} & d_{22} & d_{23} & d_{23} & d_{24} & d_{24} & d_{25} & d_{25}
\\d_{31} & d_{31} & d_{32} & d_{32} & d_{33} & d_{33} & d_{34} & d_{34} & d_{35} & d_{35}
\\d_{41} & d_{41} & d_{42} & d_{42} & d_{43} & d_{43} & d_{44} & d_{44} & d_{45} & d_{45}
\\b_{11} & b_{12} & b_{11} & b_{12} & b_{11} & b_{12} & b_{11} & b_{12} & b_{11} & b_{12}
\\b_{21} & b_{22} & b_{21} & b_{22} & b_{11} & b_{12} & b_{11} & b_{12} & b_{11} & b_{12}
\end{array}
\right)
\end{equation}
Easily we can see that the commuting condition is
\begin{eqnarray}\label{derive2}
(c_{j1}b_{i1}+c_{j1}b_{i2}+c_{j2}b_{i1}+c_{j2}b_{i2}+c_{j3}b_{i1}+c_{j3}b_{i2}+c_{j4}b_{i1}+c_{j4}b_{i2}+c_{j5}b_{i1}+c_{j5}b_{i2})\nonumber\\
+(d_{j1}a_{i1}+d_{j1}a_{i2}+d_{j2}a_{i1}+d_{j2}a_{i2}+d_{j3}a_{i1}+d_{j3}a_{i2}+d_{j4}a_{i1}+d_{j4}a_{i2}+d_{j5}a_{i1}+d_{j5}a_{i2})\nonumber\\
=(c_{j1}+c_{j2}+c_{j3}+c_{j4}+c_{j5})(b_{i1}+b_{i2})+(d_{j1}+d_{j2}+d_{j3}+d_{j4}+d_{j5})(a_{i1}+a_{i2})=0.
\end{eqnarray}
We also consider more stronger constraint conditions instead, and
different constraint conditions will give different constructions.
Actually we should not give over-optimal code too many constraint
conditions since the code may be quite difficult to be found or even
do not exist. So we consider the conditions as
$(c_1+c_2+c_3+c_4+c_5)=0$ and $(d_1+d_2+d_3+d_4+d_5)=0$ which are
both for the $[5,1,3]$ blockcode. Then by computer search we have
the blockcode:
\begin{equation}
\left(
\begin{array}{ccccc}
  $1$ & $0$ & $0$ & $1$ & $0$
\\$0$ & $1$ & $0$ & $0$ & $1$
\\$1$ & $0$ & $1$ & $0$ & $0$
\\$0$ & $1$ & $0$ & $1$ & $0$
\end{array}
\right| \left.
\begin{array}{ccccc}
  $0$ & $1$ & $1$ & $0$ & $0$
\\$0$ & $0$ & $1$ & $1$ & $0$
\\$0$ & $0$ & $0$ & $1$ & $1$
\\$1$ & $0$ & $0$ & $0$ & $1$
\end{array}
\right),
\end{equation}
and the $subcode$:
\begin{equation}
\left(
\begin{array}{cc}
  $1$ & $0$
\\$1$ & $1$
\end{array}
\right| \left.
\begin{array}{cc}
  $1$ & $1$
\\$0$ & $1$
\end{array}
\right).
\end{equation}
So the generator matrix of this $[10,4,3]$ code is
\begin{eqnarray}\label{code10}
\bordermatrix{%
& & & & & & & & & & \cr
 &X &X &Z &Z &Z &Z &X &X &I &I \cr
 &I &I &X &X &Z &Z &Z &Z &X &X \cr
 &X &X &I &I &X &X &Z &Z &Z &Z \cr
 &Z &Z &X &X &I &I &X &X &Z &Z \cr
 &Y &Z &Y &Z &Y &Z &Y &Z &Y &Z \cr
 &X &Y &X &Y &X &Y &X &Y &X &Y \cr
 }
\end{eqnarray}

\subsubsection{the optimal $[15,9,3]$, $[20,13,3]$ and $[25,18,3]$ QECC}
Similarly we can find $\{3,2\}$, $\{4,3\}$ and $\{5,3\}$
over-optimal codes which are
\begin{equation}
\left(
\begin{array}{ccc}
  $1$ & $0$ & $0$
\\$1$ & $1$ & $0$
\end{array}
\right| \left.
\begin{array}{ccc}
  $1$ & $1$ & $0$
\\$0$ & $1$ & $0$
\end{array}
\right),
\end{equation}
\begin{equation}
\left(
\begin{array}{cccc}
  $1$ & $0$ & $0$ & $1$
\\$0$ & $1$ & $0$ & $1$
\\$0$ & $0$ & $1$ & $1$
\end{array}
\right| \left.
\begin{array}{cccc}
  $0$ & $0$ & $1$ & $0$
\\$1$ & $0$ & $1$ & $1$
\\$0$ & $1$ & $0$ & $1$
\end{array}
\right),
\end{equation}
\begin{equation}
\left(
\begin{array}{ccccc}
  $1$ & $0$ & $0$ & $1$ & $1$
\\$0$ & $1$ & $0$ & $0$ & $1$
\\$0$ & $0$ & $1$ & $1$ & $1$
\end{array}
\right| \left.
\begin{array}{ccccc}
  $0$ & $0$ & $1$ & $1$ & $0$
\\$1$ & $1$ & $0$ & $1$ & $0$
\\$0$ & $1$ & $0$ & $0$ & $0$
\end{array}
\right),
\end{equation}
so the new $[15,9,3]$, $[20,13,3]$ and $[25,18,3]$ codes are:
\begin{eqnarray}\label{code15}
\bordermatrix{%
& & & & & & & & & & & & & & & \cr
 &X &X &X &Z &Z &Z &Z &Z &Z &X &X &X &I &I &I \cr
 &I &I &I &X &X &X &Z &Z &Z &Z &Z &Z &X &X &X \cr
 &X &X &X &I &I &I &X &X &X &Z &Z &Z &Z &Z &Z \cr
 &Z &Z &Z &X &X &X &I &I &I &X &X &X &Z &Z &Z \cr
 &Y &Z &I &Y &Z &I &Y &Z &I &Y &Z &I &Y &Z &I \cr
 &X &Y &I &X &Y &I &X &Y &I &X &Y &I &X &Y &I \cr
 },
\end{eqnarray}
\begin{eqnarray}\label{code20}
\bordermatrix{%
& & & & & & & & & & & & & & & & \cr
 &X &X &X &X  &I &I &I &I  &Z &Z &Z &Z  &Z &Z &Z &Z  &Y &Y &Y &Y  \cr
 &Z &Z &Z &Z  &X &X &X &X  &I &I &I &I  &Z &Z &Z &Z  &X &X &X &X  \cr
 &Z &Z &Z &Z  &Z &Z &Z &Z  &Y &Y &Y &Y  &I &I &I &I  &Y &Y &Y &Y  \cr
 &I &I &I &I  &Z &Z &Z &Z  &Z &Z &Z &Z  &Y &Y &Y &Y  &X &X &X &X  \cr
 &X &I &Z &X  &X &I &Z &X  &X &I &Z &X  &X &I &Z &X  &X &I &Z &X  \cr
 &Z &X &Z &Y  &Z &X &Z &Y  &Z &X &Z &Y  &Z &X &Z &Y  &Z &X &Z &Y  \cr
 &I &Z &X &Y  &I &Z &X &Y  &I &Z &X &Y  &I &Z &X &Y  &I &Z &X &Y  \cr
 },
\end{eqnarray}
\begin{eqnarray}\label{code2518}
\bordermatrix{%
& & & & & & & & & & & & & & & & \cr
 &X &X &X &X &X &Z &Z &Z &Z &Z &Z &Z &Z &Z &Z &X &X &X &X &X &I &I &I &I &I \cr
 &I &I &I &I &I &X &X &X &X &X &Z &Z &Z &Z &Z &Z &Z &Z &Z &Z &X &X &X &X &X \cr
 &X &X &X &X &X &I &I &I &I &I &X &X &X &X &X &Z &Z &Z &Z &Z &Z &Z &Z &Z &Z \cr
 &Z &Z &Z &Z &Z &X &X &X &X &X &I &I &I &I &I &X &X &X &X &X &Z &Z &Z &Z &Z \cr
 &X &I &Z &Z &X &X &I &Z &Z &X &X &I &Z &Z &X &X &I &Z &Z &X &X &I &Z &Z &X \cr
 &Z &Y &I &Z &X &Z &Y &I &Z &X &Z &Y &I &Z &X &Z &Y &I &Z &X &Z &Y &I &Z &X \cr
 &I &Z &X &X &X &I &Z &X &X &X &I &Z &X &X &X &I &Z &X &X &X &I &Z &X &X &X \cr
 }.
\end{eqnarray}

\subsubsection{Another construction of $[10,4,3]$ code}
Here we give another way of constructing $[10,4,3]$ code. We may
notice that two copies of a $[5,1,3]$ code whose syndromes are
$\{\{S'_x\},\{S'_z\},\{S'_y\}\}$ could build a $\{10,4\}$
over-optimal code in this way:
\begin{eqnarray}
\{\{S_x\}=\{S'_x\}\cup\{S'_z\},\nonumber\\
\{S_z\}=\{S'_z\}\cup\{S'_y\},\nonumber\\
\{S_y\}=\{S'_y\}\cup\{S'_x\}\},
\end{eqnarray}
then we choose $\bordermatrix{%
& & \cr
 &1 &0\cr
 &0 &1\cr}$
as blockcode, so commuting conditions of blockcode and subcode are
satisfied automatically. Commuting conditions between blockcode and
subcode is $a_1(d_1+\cdots+d_{10})+b_1(c_1+\cdots+c_{10})=0$, which
means the number of "1"s in any row of both halves of generator
matrices of $\{10,4\}$ code is even. Instead we turn to search a
$[5,1,3]$ code which satisfies the same condition, then it is easy
to prove that the constructed $\{10,4\}$ code is appropriate. It is
obvious that the $[5,1,3]$ code above
\begin{equation}
\left(
\begin{array}{ccccc}
  $1$ & $0$ & $0$ & $1$ & $0$
\\$0$ & $1$ & $0$ & $0$ & $1$
\\$1$ & $0$ & $1$ & $0$ & $0$
\\$0$ & $1$ & $0$ & $1$ & $0$
\end{array}
\right| \left.
\begin{array}{ccccc}
  $0$ & $1$ & $1$ & $0$ & $0$
\\$0$ & $0$ & $1$ & $1$ & $0$
\\$0$ & $0$ & $0$ & $1$ & $1$
\\$1$ & $0$ & $0$ & $0$ & $1$
\end{array}
\right)
\end{equation}
is feat. Then the $[10,4,3]$ code is
\begin{eqnarray}\label{code102}
\bordermatrix{%
& & & & & & & & & & \cr
 &X &X &X &X &X &X &X &X &X &X \cr
 &Z &Z &Z &Z &Z &Z &Z &Z &Z &Z \cr
 &X &Z &Z &X &I &Z &Y &Y &Z &I \cr
 &I &X &Z &Z &X &I &Z &Y &Y &Z \cr
 &X &I &X &Z &Z &Z &I &Z &Y &Y \cr
 &Z &X &I &X &Z &Y &Z &I &Z &Y \cr
 }.
\end{eqnarray}

\subsection{Gottesman's $[2^k,2^k-k-2,3]$ codes and perfect codes}

The over-optimal codes whose $S_x$, $S_z$ and $S_y$ syndromes are
different themselves are very important codes in our method of
constructing optimal codes, so we call them $optimal-constructing$
codes. More important thing is to know $g$'s upper and lower bounds
of optimal-constructing codes and when the codes could reach the
upper bound. Here we have the lemma below:

$Lemma$: For a $[n,n-k,d]$ optimal-constructing code,
$1<g\leq\frac{3(2^k-1)}{2^k}<3$. When $n=2^k-1$, the code could
reach the upper bound.

$Proof$: A $[n,n-k,d]$ code could provide $2^k$ syndromes. If each
of $S_x$, $S_z$ and $S_y$ takes all the syndromes except for
$00\cdots0$ and $S^x+S^z=S^y$, this code must reach the upper bound.
So $g$'s upper bound is $\frac{3(2^k-1)}{2^k}$. But not
optimal-constructing codes of all length could reach upper bound,
since each of $S_x$, $S_z$ and $S_y$ takes all the syndromes except
for $00\cdots0$, $n$ must equal to $2^k-1$.

So $[2^k-1,2^k-1-k,<3]$ codes are optimal-constructing codes that
reach the upper bound, and we call them $raw-perfect-constructing$
codes. Since $\lim_{k\rightarrow\infty}\frac{3(2^k-1)}{2^k}=3$,
these code are quite useful to constructing optimal codes and
perfect codes. And for the raw perfect-constructing codes we have
the lemma below:

$Lemma$: Suppose any generator of a $[2^k-1,2^k-1-k,<3]$ code is
$(c_1,\cdots,c_{2^k-1}|d_1,\cdots,d_{2^k-1})$, then
$(c_1+\cdots+c_{2^k-1})=0$ and $(d_1+\cdots+d_{2^k-1})=0$

$Proof$: Since this code used all syndromes that the code could
provide, any row of the left half of generator matrix will have
$2^{k-1}$ "1"s which means the number of "1"s is always even, so
$(c_1+\cdots+c_{2^k-1})=0$. The same to the right half which means
$(d_1+\cdots+d_{2^k-1})=0$.

1. Gottesman's $[2^k,2^k-k-2,3]$ codes are one of this kind. A
$[2^k,2^k-k-2,3]$ code of this class can be constructed by nesting a
"code" which is $\left(
\begin{array}{c}
  $1$
\\$0$
\end{array}
\right| \left.
\begin{array}{c}
  $0$
\\$1$
\end{array}
\right)$ and a $[2^k-1,2^k-1-k,<3]$ raw perfect-constructing code
together. $\left(
\begin{array}{c}
  $1$
\\$0$
\end{array}
\right| \left.
\begin{array}{c}
  $0$
\\$1$
\end{array}
\right)$ is blockcode whose syndromes are $\{10,01,11\}$
 and the raw perfect-constructing code is subcode whose syndromes
 are:$\{\{S_x\},\{S_z\},\{S_y\}\}$. The code is constructed this
 form:
\begin{eqnarray}
\{\{10\}\otimes\{S_x,00\cdots0\},\nonumber\\
\{01\}\otimes\{S_z,00\cdots0\},\nonumber\\
\{11\}\otimes\{S_y,00\cdots0\}\}
\end{eqnarray}
So $g$ of this class is always
$\frac{1}{3}\cdot\frac{3}{4}\cdot3=\frac{3}{4}$.

 Here we show how to construct the second code of this
 class--
$[16,10,3]$. Firstly we need to construct $\{15,4\}$ raw
perfect-constructing code. We could choose any $[5,1,3]$ code whose
syndromes are $\{\{S'_x\},\{S'_z\},\{S'_y\}\}$, and the $\{15,4\}$
code takes this form:
\begin{eqnarray}
\{\{S_x\}=\{S'_x\}\cup\{S'_z\}\cup\{S'_y\},\nonumber\\
\{S_z\}=\{S'_z\}\cup\{S'_y\}\cup\{S'_x\},\nonumber\\
\{S_y\}=\{S'_y\}\cup\{S'_x\}\cup\{S'_z\}\}
\end{eqnarray}
It is easy to prove that $S_x^i+S_z^i=S_y^i$ and the commute
conditions of subcode and blockcode are satisfied automatically.
Commute condition between subcode and blockcode is:
$a_1(d_1+\cdots+d_{16})+b_1(c_1+\cdots+c_{16})=0$, and according to
the lemma above this condition is also satisfied automatically. So
if we choose a $[5,1,3]$ code as
\begin{eqnarray}
\bordermatrix{%
& & & & & \cr
 &X &I &Z &Z &Y \cr
 &Z &X &I &Z &X \cr
 &Z &Z &Y &I &Y \cr
 &I &Z &Z &Y &X \cr
 },
\end{eqnarray}
then the $[16,10,3]$ code is
\begin{eqnarray}\label{code16}
\bordermatrix{%
& & & & & & & & & & & & & & & & \cr
 &X &X &X &X &X &X &X &X &X &X &X &X &X &X &X &X \cr
 &Z &Z &Z &Z &Z &Z &Z &Z &Z &Z &Z &Z &Z &Z &Z &Z \cr
 &X &I &Z &Z &Y &Z &I &Y &Y &X &Y &I &X &X &Z &I \cr
 &Z &X &I &Z &X &Y &Z &I &Y &Z &X &Y &I &X &Y &I \cr
 &Z &Z &Y &I &Y &Y &Y &X &I &X &X &X &Z &I &Z &I \cr
 &I &Z &Z &Y &X &I &Y &Y &X &Z &I &X &X &Z &Y &I \cr
 }
\end{eqnarray}

$[8,3,3]$ code have been constructed which is
\begin{equation}
\left(
\begin{array}{cccccccc}
  1&1&1&1&1&1&1&1
\\0&0&0&0&0&0&0&0
\\0&0&0&0&1&1&1&1
\\0&0&1&1&0&0&1&1
\\0&1&0&1&0&1&0&1
\end{array}
\right| \left.
\begin{array}{cccccccc}
  0&0&0&0&0&0&0&0
\\1&1&1&1&1&1&1&1
\\0&1&0&1&1&0&1&0
\\0&1&0&1&0&1&0&1
\\0&1&1&0&1&0&0&1
\end{array}
\right)
\end{equation}
But here we give another construction of $[8,3,3]$ that out of the
method of building $[2^k,2^k-k-2,3]$ codes, which is:
\begin{equation}
\left(
\begin{array}{cccccccc}
  1&1&1&1&0&0&0&0
\\0&0&0&0&1&1&1&1
\\0&0&0&0&1&1&1&1
\\0&0&1&1&0&0&1&1
\\0&1&0&1&0&1&0&1
\end{array}
\right| \left.
\begin{array}{cccccccc}
  0&1&0&1&1&0&1&0
\\1&0&1&0&0&1&0&1
\\0&1&0&1&1&0&1&0
\\0&1&0&1&0&1&0&1
\\0&1&1&0&1&0&0&1
\end{array}
\right)
\end{equation}

2. The perfect one-error-correcting codes are also one of this kind.
This class is $[(2^k-1)/3,(2^k-1)/3-k,3]$ codes whose $k$ must be
even. Though the first one of this class which is $\bordermatrix{%
& & \cr
 &1 &0\cr
 &0 &1\cr}$ is not a real code, we still call it "perfect code" since it is quite important in code construction. Comparing with
other methods of constructing perfect codes, our method is more
general. We have the theory below:

$Theory$: A $[(2^{k+k'}-1)/3,(2^{k+k'}-1)/3-(k+k'),3]$ perfect code
can be constructed by nesting a $[(2^k-1)/3,(2^k-1)/3-k,3]$ perfect
code, a $[2^{k'}-1,2^{k'}-1-{k'},<3]$ raw perfect-constructing code
and a $[(2^{k'}-1)/3,(2^{k'}-1)/3-{k'},3]$ perfect code together.

$Proof$: Firstly we nesting a $[(2^k-1)/3,(2^k-1)/3-k,3]$ perfect
code and a $[2^{k'}-1,2^{k'}-1-{k'},<3]$ raw perfect-constructing
code together in this way:
\begin{eqnarray}
\{\{B_x\}\otimes\{S_x,00\cdots0\},\nonumber\\
\{B_z\}\otimes\{S_z,00\cdots0\},\nonumber\\
\{B_y\}\otimes\{S_y,00\cdots0\}\}
\end{eqnarray}
Since all commute conditions are satisfied automatically, this
construct a $[(2^{k+k'}-2^{k'})/3,(2^{k+k'}-2^{k'})/3-(k+k'),3]$
optimal code. Notice that we never used syndrome $00\cdots0$ of
blockcode, and the  raw perfect-constructing code has $k'$
generators,  so we could use syndrome $00\cdots0$ to nest another
perfect code which also has $k'$ generators. This
$[(2^{k'}-1)/3,(2^{k'}-1)/3-{k'},3]$ perfect code is another subcode
with syndromes $\{S'_x,S'_z,S'_y\}$. Finally this
$[(2^{k+k'}-1)/3,(2^{k+k'}-1)/3-(k+k'),3]$ code is constructed:
\begin{eqnarray}
\{\{B_x\}\otimes\{S_x,00\cdots0\},\{B_z\}\otimes\{S_z,00\cdots0\},\nonumber\\
\{B_y\}\otimes\{S_y,00\cdots0\},00\cdots0\otimes\{S'_x\},\nonumber\\
00\cdots0\otimes\{S'_z\},00\cdots0\otimes\{S'_y\}\}
\end{eqnarray}
It is easy too prove that all commute conditions are satisfied.

In process of code construction, $g$ of the
$[(2^{k+k'}-2^{k'})/3,(2^{k+k'}-2^{k'})/3-(k+k'),3]$ optimal code is
$g_1=\frac{1}{3}\cdot(1-\frac{1}{2^k})\cdot3=1-\frac{1}{2^k}$ and
$g$ of $00\cdots0$ pasted with $[(2^{k'}-1)/3,(2^{k'}-1)/3-{k'},3]$
code is $g_2=\frac{1}{2^k}\cdot(1-\frac{1}{2^{k'}})$, so for the
$[(2^{k+k'}-1)/3,(2^{k+k'}-1)/3-(k+k'),3]$ code
$g=g_1+g_2=1-\frac{1}{2^{k+k'}}$. Obviously it is a perfect code.

Actually our method is not constrained by whether $k$ and $k'$ are
even or odd, which is mentioned in Ref.\cite{CRSS1}. For example,
$[40,33,3]$ code can also be constructed using our method.

\subsection{Gottesman's stabilizer pasting}

$Stabilizier pasting$: Given two nondegenerate stabilizer codes
$[n_2,s_2]=\langle S_1,\cdots,S_{s_2}\rangle$ and $[n_1,s_1]=\langle
T_1,\cdots,T_{s_1}\rangle$ of distance 3, if two observables
$X(n_2)$ and $Z(n_2)$ belong to $[n_2,s_2]$, say, $S_1=X(n_2)$ and
$S_2=Z(n_2)$, then the stabilizer constructed in the way below
defines a nondegenerate stabilizer code $[n_2+n_1,s]$ with
$s=max\{s_2,s_1+2\}$, denoted as $[n_2,s_2]\rhd[n_1,s_1]$.

Consider his theory with our method, the blockcode of $[n_2,s_2]$ is always $\bordermatrix{%
& & \cr
 &1 &0\cr
 &0 &1\cr}$ we mentioned above, and the subcode may not be a real code since commute
conditions may not be satisfied. Notice that syndrome 00 of
blockcode has never been used, so we put another code $[n_1,s_1]$
under 00 to finish the construction. Code $[n_1,s_1]$ may also not
be a real code as long as the final constructed code satisfies all
commute conditions. Many optimal codes have been constructed in Ref.
\cite{yu}.

To construct the $[37,30,3]$ code, firstly we choose the nontrivial
$[8,3,3]$ code found above
\begin{equation}
\left(
\begin{array}{cccccccc}
  1&1&1&1&0&0&0&0
\\0&0&0&0&1&1&1&1
\\0&0&0&0&1&1&1&1
\\0&0&1&1&0&0&1&1
\\0&1&0&1&0&1&0&1
\end{array}
\right| \left.
\begin{array}{cccccccc}
  0&1&0&1&1&0&1&0
\\1&0&1&0&0&1&0&1
\\0&1&0&1&1&0&1&0
\\0&1&0&1&0&1&0&1
\\0&1&1&0&1&0&0&1
\end{array}
\right)
\end{equation}
as blockcode and a $\{4,2\}$ code found by computer search
\begin{equation}
\left(
\begin{array}{cccc}
  1&0&0&1
\\1&1&0&0
\end{array}
\right| \left.
\begin{array}{cccc}
  1&1&0&0
\\1&0&1&0
\end{array}
\right)
\end{equation}
to be subcode. There have 8 syndromes that the $[8,3,3]$ code did
not used which are:
\begin{eqnarray}
\bordermatrix{%
& & & & & & & & \cr
 &0 &1 &0 &1 &0 &1 &0 &1 \cr
 &0 &1 &0 &1 &0 &1 &0 &1 \cr
 &0 &1 &0 &1 &0 &1 &0 &1 \cr
 &0 &1 &1 &0 &0 &1 &1 &0 \cr
 &0 &0 &1 &1 &1 &1 &0 &0 \cr
 }.
\end{eqnarray}
Since syndromes of $[37,30,3]$ code are 7-bit numbers, with another
two bits we can find a $\{5,7\}$ code below:
\begin{equation}
\left(
\begin{array}{ccccc}
  0&1&0&1&0
\\0&1&0&1&0
\\0&1&0&1&0
\\1&0&0&1&0
\\0&1&0&0&1
\\1&0&1&0&0
\\0&1&0&1&0
\end{array}
\right| \left.
\begin{array}{ccccc}
  1&0&0&0&1
\\1&0&0&0&1
\\1&0&0&0&1
\\0&1&1&0&0
\\0&0&1&1&0
\\0&0&0&1&1
\\1&0&0&0&1
\end{array}
\right).
\end{equation}
The last four generators build a $[5,1,3]$ code and the syndromes of
first five generators all belong to those 8 syndromes mentioned
above. It is easy to check that all commute conditions are
satisfied, then the $[37,30,3]$ code is constructed as

{\footnotesize
\begin{equation}\label{code37}
\left(
\begin{array}{ccccccccccccccccccccccccccccccccccccc}
  X &X &X &X &Y &Y &Y &Y &X &X &X &X &Y
  &Y &Y &Y &Z &Z &Z &Z &I &I &I &I &Z &Z &Z &Z &I &I &I &I &Z &X &I &X &Z
\\Z &Z &Z &Z &I &I &I &I &Z &Z &Z &Z &I &I &I &I &X &X &X &X &Y &Y &Y &Y &X &X &X &X &Y &Y &Y &Y &Z &X &I &X &Z
\\I &I &I &I &Z &Z &Z &Z &I &I &I &I &Z &Z &Z &Z &Y &Y &Y &Y &X &X &X &X &Y &Y &Y &Y &X &X &X &X &Z &X &I &X &Z
\\I &I &I &I &Z &Z &Z &Z &X &X &X &X &Y &Y &Y &Y &I &I &I &I &Z &Z &Z &Z &X &X &X &X &Y &Y &Y &Y &X &Z &Z &X &I
\\I &I &I &I &Y &Y &Y &Y &Z &Z &Z &Z &X &X &X &X &Z &Z &Z &Z &X &X &X &X &I &I &I &I &Y &Y &Y &Y &I &X &Z &Z &X
\\Y &Z &I &Z &Y &Z &I &Z &Y &Z &I &Z &Y &Z &I &Z &Y &Z &I &Z &Y &Z &I &Z &Y &Z &I &Z &Y &Z &I &Z &X &I &X &Z &Z
\\Y &X &Z &I &Y &X &Z &I &Y &X &Z &I &Y &X &Z &I &Y &X &Z &I &Y &X &Z &I &Y &X &Z &I &Y &X &Z &I &Z &X &I &X &Z
\end{array}
\right).
\end{equation}
}

\section{Codes Construction for all Distance}

Suppose there are two copies of one $[5,1,3]$ code whose generators
and logical operators are $M_1,M_2,M_3,M_4$ and $\bar{X},\bar{Z}$
pasted together. If the construction is $\bordermatrix{%
& & \cr
 &A &0\cr
 &0 &A\cr}$, then the logical operators of this code are $\{\bar{X}\otimes I,I\otimes\bar{X},\bar{Z}\otimes
 I,I\otimes\bar{Z}\}$; if the construction is $\bordermatrix{%
& & \cr
 &A &A\cr}$, then the logical operators of this code are $\{\bar{X}\otimes I,I\otimes\bar{X},\bar{Z}\otimes
 I,I\otimes\bar{Z},M_1\otimes I,M_2\otimes I,M_3\otimes I,M_4\otimes
 I,\bar{M_1}\otimes\bar{M_1},\bar{M_2}\otimes\bar{M_2},\bar{M_3}\otimes\bar{M_3},\bar{M_4}\otimes\bar{M_4}\}$.
 Obviously the first one is the easiest way to construct good
 blockcode and the second one is the hardest.

\subsection{Concatenated coding}

Suppose 5 copies of a $[5,1,3]$ code are pasted in the easiest way,
then the constructed $[25,20]$ code has 20 generators and 10 logical
operators. So if we want to build code of larger distance, we must
pick some logical operators and put them into the stabilizer, and
these logical operators build the $blockcode$. Suppose the logical X
and Z operators of the five copies are
$\bar{X_1},\bar{X_2},\bar{X_3},\bar{X_4},\bar{X_5}$ and
$\bar{Z_1},\bar{Z_2},\bar{Z_3},\bar{Z_4},\bar{Z_5}$, it is easy to
think out that a 5-qubit code whose physical qubits are replace by
these ten logical operators can be constructed. We choose this
5-qubit code to be $blockcode$, and if distance of $blockcode$ is
$d$ then distance of this constructed code will be $3d$. As we know
the largest distance of 5-qubit codes is 3, so choose any $[5,1,3]$
code to be $blockcode$ we will get a $[25,24,9]$ code. For example,
if we choose
\begin{equation}
\left(
\begin{array}{ccccc}
  $1$ & $0$ & $0$ & $1$ & $0$
\\$0$ & $1$ & $0$ & $0$ & $1$
\\$1$ & $0$ & $1$ & $0$ & $0$
\\$0$ & $1$ & $0$ & $1$ & $0$
\end{array}
\right| \left.
\begin{array}{ccccc}
  $0$ & $1$ & $1$ & $0$ & $0$
\\$0$ & $0$ & $1$ & $1$ & $0$
\\$0$ & $0$ & $0$ & $1$ & $1$
\\$1$ & $0$ & $0$ & $0$ & $1$
\end{array}
\right),
\end{equation}
then the generator matrix is
\begin{eqnarray}\label{concatenatedcode}
\bordermatrix{%
& & & & & & & & & & & & & & & & \cr
 &X &X &X &X &X &Z &Z &Z &Z &Z &Z &Z &Z &Z &Z &X &X &X &X &X &I &I &I &I &I \cr
 &I &I &I &I &I &X &X &X &X &X &Z &Z &Z &Z &Z &Z &Z &Z &Z &Z &X &X &X &X &X \cr
 &X &X &X &X &X &I &I &I &I &I &X &X &X &X &X &Z &Z &Z &Z &Z &Z &Z &Z &Z &Z \cr
 &Z &Z &Z &Z &Z &X &X &X &X &X &I &I &I &I &I &X &X &X &X &X &Z &Z &Z &Z &Z \cr
 &X &Z &Z &X &I &I &I &I &I &I &I &I &I &I &I &I &I &I &I &I &I &I &I &I &I \cr
 &I &X &Z &Z &X &I &I &I &I &I &I &I &I &I &I &I &I &I &I &I &I &I &I &I &I \cr
 &X &I &X &Z &Z &I &I &I &I &I &I &I &I &I &I &I &I &I &I &I &I &I &I &I &I \cr
 &Z &X &I &X &Z &I &I &I &I &I &I &I &I &I &I &I &I &I &I &I &I &I &I &I &I \cr
 &I &I &I &I &I &X &Z &Z &X &I &I &I &I &I &I &I &I &I &I &I &I &I &I &I &I \cr
 &I &I &I &I &I &I &X &Z &Z &X &I &I &I &I &I &I &I &I &I &I &I &I &I &I &I \cr
 &I &I &I &I &I &X &I &X &Z &Z &I &I &I &I &I &I &I &I &I &I &I &I &I &I &I \cr
 &I &I &I &I &I &Z &X &I &X &Z &I &I &I &I &I &I &I &I &I &I &I &I &I &I &I \cr
 &I &I &I &I &I &I &I &I &I &I &X &Z &Z &X &I &I &I &I &I &I &I &I &I &I &I \cr
 &I &I &I &I &I &I &I &I &I &I &I &X &Z &Z &X &I &I &I &I &I &I &I &I &I &I \cr
 &I &I &I &I &I &I &I &I &I &I &X &I &X &Z &Z &I &I &I &I &I &I &I &I &I &I \cr
 &I &I &I &I &I &I &I &I &I &I &Z &X &I &X &Z &I &I &I &I &I &I &I &I &I &I \cr
 &I &I &I &I &I &I &I &I &I &I &I &I &I &I &I &X &Z &Z &X &I &I &I &I &I &I \cr
 &I &I &I &I &I &I &I &I &I &I &I &I &I &I &I &I &X &Z &Z &X &I &I &I &I &I \cr
 &I &I &I &I &I &I &I &I &I &I &I &I &I &I &I &X &I &X &Z &Z &I &I &I &I &I \cr
 &I &I &I &I &I &I &I &I &I &I &I &I &I &I &I &Z &X &I &X &Z &I &I &I &I &I \cr
 &I &I &I &I &I &I &I &I &I &I &I &I &I &I &I &I &I &I &I &I &X &Z &Z &X &I \cr
 &I &I &I &I &I &I &I &I &I &I &I &I &I &I &I &I &I &I &I &I &I &X &Z &Z &X \cr
 &I &I &I &I &I &I &I &I &I &I &I &I &I &I &I &I &I &I &I &I &X &I &X &Z &Z \cr
 &I &I &I &I &I &I &I &I &I &I &I &I &I &I &I &I &I &I &I &I &Z &X &I &X &Z \cr
 }.
\end{eqnarray}

This is so-called concatenated coding which has already been pointed
by Gottesman in Ref.\cite{gottesman0}. The codes constructed in this
way must be degenerate code since distance only increases in
$N(S)-S$ but distance of $S$ does not change. Here we see that the
concatenated coding construction can be understood in the framework
of the nested coding construction.

\subsection{Stabilizer pasting for all distance}

Theory of general stabilizer pasting for any distance is:

$Theory$: Suppose there have four stabilizers $R_1$, $R_2$, $S_1$,
and $S_2$ with $R_1\subset S_1$ and $R_2\subset S_2$. Let $R_1$
 define a $[n_1,l_1,c_1]$ code, $R_1$ be a $[n_2,l_2,c_2]$ code, $S_1$
 be a $[n_1,k_1,d_1]$ code and $S_2$ be a $[n_2,k_2,d_2]$ code with
 $k_i<l_i$, $c_i\leq d_i$, $l_1-k_1=l_2-k_2$ and $S_1$, $S_2$ to be
 nondegenerate. Let generators of $R_1$ be
 $\{M_1,\cdots,M_{n_1-l_1}\}$, generators of $S_1$ be
 $\{M_1,\cdots,M_{n_1-k_1}\}$, generators of $R_2$ be
 $\{N_1,\cdots,N_{n_2-l_2}\}$, and generators of $S_2$ be
 $\{N_1,\cdots,N_{n_2-k_2}\}$. Then we can build a new stabilizer
 $S$ on $n_1+n_2$ qubits generated by
\begin{eqnarray}
\{M_1\otimes I,\cdots,M_{n_1-l_1}\otimes I,I\otimes
N_1,\cdots,I\otimes N_{n_2-l_2},\nonumber\\
M_{n_1-l_1+1}\otimes N_{n_2-l_2+1},\cdots,M_{n_1-k_1}\otimes
N_{n_2-k_2}\}.
\end{eqnarray}
The code has $(n_1-l_1)+(n_2-l_2)+(l_i-k_i)$ generators which means
it encodes $l_1+k_2=l_2+k_1$ qubits, and the distance of the new
code is $min\{d_1,d_2,c_1+c_2\}$.

Since $R_1$ and $R_2$ are connected in block diagonal matrix,
obviously they are subcodes and the $blockcode$ is built by all
logical operators of them. Suppose logical operators of $R_1$ are
$\{\bar{X_1},\cdots,\bar{X_{l_1}},\bar{Z_1},\cdots,\bar{Z_{l_1}}\}$,
and logical operators of $R_2$ are
$\{\bar{X'_1},\cdots,\bar{X'_{l_2}},\bar{Z'_1},\cdots,\bar{Z'_{l_2}}\}$.
Then we can rewrite $S_1$ as
$\{M_1,\cdots,M_{n_1-l_1},\bar{X_1},\cdots,\bar{X_{l_1-k_1}}\}$ and
$S_2$ as
$\{N_1,\cdots,N_{n_2-l_2},\bar{X'_1},\cdots,\bar{X'_{l_2-k_2}}\}$,
so logical operators of $S_1$ and $S_2$ are
$\{\bar{X_{l_1-k_1+1}},\cdots,\bar{X_{l_1}},\bar{Z_{l_1-k_1+1}},\cdots,\bar{Z_{l_1}}\}$
and
$\{\bar{X'_{l_2-k_2+1}},\cdots,\bar{X'_{l_2}},\bar{Z'_{l_2-k_2+1}},\cdots,\bar{Z'_{l_2}}\}$.
Obviously the blockcode is
$\{\bar{X_1}\otimes\bar{X'_1},\cdots,\bar{X_{l_1-k_1}}\otimes\bar{X'_{l_2-k_2}}\}$
and the logical operators of the constructed code are
$\{\bar{X_{l_1-k_1+1}}\otimes I,\cdots,\bar{X_{l_1}}\otimes
I,\bar{Z_{l_1-k_1+1}}\otimes I,\cdots,\bar{Z_{l_1}}\otimes
I\}\cup\{I\otimes\bar{X'_{l_2-k_2+1}},\cdots,I\otimes\bar{X'_{l_2}},I\otimes\bar{Z'_{l_2-k_2+1}},\cdots,I\otimes\bar{Z'_{l_2}}\}\cup\{\bar{X_1}\otimes
I,\cdots,\bar{X_{l_1-k_1}}\otimes
I\}\cup\{\bar{Z_1}\otimes\bar{Z'_1},\cdots,\bar{Z_{l_1-k_1}}\otimes\bar{Z'_{l_2-k_2}}\}$.

We can see that there are three kinds of logical operators. The
first one are logical operators of $S_1$ and $S_2$, which act on
stabilizer and make the minimal weight be $min\{d_1,d_2\}$. The
second one are $\{\bar{X_1}\otimes I,\cdots,\bar{X_{l_1-k_1}}\otimes
I\}$, and since $S_1$ and $S_2$ are nondegenerate, they act on
stabilizer and make the minimal weight be $min\{d_1,d_2\}$. The last
one are
$\{\bar{Z_1}\otimes\bar{Z'_1},\cdots,\bar{Z_{l_1-k_1}}\otimes\bar{Z'_{l_2-k_2}}\}$,
which act on stabilizer and make the minimal weight be $c_1+c_2$. So
the distance of the new constructed code is
$min\{d_1,d_2,c_1+c_2\}$.

\end{document}